\documentclass[sigconf,natbib=true,anonymous=false]{acmart} % resource papers are single blind 

\AtBeginDocument{%
    \providecommand\BibTeX{{%
            \normalfont B\kern-0.5em{\scshape i\kern-0.25em b}\kern-0.8em\TeX}}}

\usepackage[T1]{fontenc}
\usepackage[latin1]{inputenc}
\usepackage{array, amsmath}
\usepackage{xspace}
\usepackage{xcolor}
\usepackage{url}% [hyphens] option leads to option clash.

\appto\UrlBreaks{\do\-}

\newcommand{\Ni}{(1)~}
\newcommand{\Nii}{(2)~}

\usepackage{tcolorbox}

\newcommand{\textttsmall}[1]{\texttt{\small #1}}

\newcommand{\irdatasets}{\textttsmall{ir\_datasets}\xspace}
\newcommand{\irmeasures}{\textttsmall{ir\_measures}\xspace}
\newcommand{\irmetadata}{\textttsmall{ir\_metadata}\xspace}
\newcommand{\reproeval}{\textttsmall{repro\_eval}\xspace}

\DeclareGraphicsExtensions{.ai,.pdf,.jpg,.png}
\setkeys{Gin}{pagebox=artbox}% Alternative (necessary for newer tex versions) for \pdfpagebox 5 in preceding line.
\graphicspath{{.}}
\newcommand{\bsfigure}[3][scale=1.0]{%
  \begin{figure}[tb]
    \centering
    \includegraphics[#1]{#2}
    \vspace{-4ex}
    \caption{#3}\label{#2}
    \Description{#3}
    \vspace{-2ex}
  \end{figure}}

\usepackage{float}
\usepackage{listings}
\usepackage{listingsutf8}
\definecolor{mediumgray}{gray}{0.60}
\definecolor{webiscodebasic}{rgb}{0.2,0.2,0.2}
\definecolor{webiscodekeyword}{rgb}{0.0,0.5,0.0}
\definecolor{webiscodekeywordself}{rgb}{0.7,0.4,0.6}
\definecolor{webiscodeidentifier}{rgb}{0.0,0.0,0.0}
\definecolor{webiscodecomment}{rgb}{0.25,0.5,0.5}
\definecolor{webiscodestring}{rgb}{0.75,0.12,0.12}
\definecolor{webiscodedecorator}{rgb}{0.6,0.3,0.0}
\lstdefinestyle{webisstyle}{
  basicstyle=\fontsize{7.45pt}{8.45pt}\selectfont\ttfamily\color{webiscodebasic},
  keywordstyle=\color{webiscodekeyword},
  identifierstyle=\color{webiscodeidentifier},
  commentstyle=\color{webiscodecomment},
  stringstyle=\color{webiscodestring},
  showstringspaces=false,
  frame=lines,
  framesep=0.7em,
  rulesep=0.5em,
  framerule=0.1em,
  keepspaces=true,
  tabsize=2,
  showtabs=false,
  numbers=none,
  literate=
    {->}{{\textrightarrow}}{2}
    {>=}{{\(\geq\)}}{2}
    {<=}{{\(\leq\)}}{2}
    {!=}{{\(\neq\)}}{2},
}
\newfloat{listing}{htbp}{lop}
\floatname{listing}{Listing}

\usepackage{multirow}

\newif\ifbscomment
\bscommentfalse
\bscommenttrue% Show comments.

\RequirePackage{color}
\RequirePackage{soul}
\setstcolor{blue}
\definecolor{linkblue}{rgb}{0.0, 0.33, 0.88}

\usepackage{pifont}
\newcommand{\cmark}{\ding{51}}%
\newcommand{\xmark}{\ding{55}}%

\begin{document}
    
\title{The Information Retrieval Experiment Platform}

\author{Maik Fr{\"o}be}
\affiliation{
\institution{Friedrich-Schiller-Universit{\"a}t Jena}
\country{}
\city{}
}

\author{Jan Heinrich Reimer}
\affiliation{
\institution{Friedrich-Schiller-Universit{\"a}t Jena}
\country{}
\city{}
}

\author{Sean MacAvaney}
\affiliation{
\institution{University of Glasgow}
\country{}
\city{}
}

\author{Niklas Deckers}
\affiliation{
\institution{Leipzig University and ScaDS.AI}
\country{}
\city{}
}

\author{Simon Reich}
\affiliation{
\institution{Leipzig University}
\country{}
\city{}
}

\author{Janek Bevendorff}
\affiliation{
\institution{Bauhaus-Universit{\"a}t Weimar}
\country{}
\city{}
}

\author{Benno Stein}
\affiliation{
\institution{Bauhaus-Universit{\"a}t Weimar}
\country{}
\city{}
}

\author{Matthias Hagen}
\affiliation{
\institution{Friedrich-Schiller-Universit{\"a}t Jena}
\country{}
\city{}
}

\author{Martin Potthast}
\affiliation{
\institution{Leipzig University and ScaDS.AI}
\country{}
\city{}
}

\renewcommand{\shortauthors}{Maik Fr{\"o}be et al.}

\begin{abstract}
We integrate \irdatasets, \irmeasures, and PyTerrier with TIRA in the Information Retrieval Experiment Platform~(TIREx) to promote more standardized, reproducible, scalable, and even blinded retrieval experiments. Standardization is achieved when a retrieval approach implements PyTerrier's interfaces and the input and output of an experiment are compatible with \irdatasets and \irmeasures. However, none of this is a must for reproducibility and scalability, as TIRA can run any dockerized software locally or remotely in a cloud-native execution environment. Version control and caching ensure efficient (re)execution. TIRA allows for blind evaluation when an experiment runs on a remote server or cloud not under the control of the experimenter. The test data and ground truth are then hidden from public access, and the retrieval software has to process them in a sandbox that prevents data leaks.

We currently host an instance of TIREx with 15~corpora (1.9~billion documents) on which 32~shared retrieval tasks are based. Using Docker images of 50~standard retrieval approaches, we automatically evaluated all approaches on all tasks (50 $\cdot$ 32 = 1,600~runs) in less than a week on a midsize cluster (1,620~CPU cores and 24~GPUs). This instance of TIREx is open for submissions and will be integrated with the IR~Anthology, as well as released open source.
\end{abstract}

\copyrightyear{2023}
\acmYear{2023}
\setcopyright{acmlicensed}\acmConference[SIGIR '23]{Proceedings of the 46th International ACM SIGIR Conference on Research and Development in Information Retrieval}{July 23--27, 2023}{Taipei, Taiwan}
\acmBooktitle{Proceedings of the 46th International ACM SIGIR Conference on Research and Development in Information Retrieval (SIGIR '23), July 23--27, 2023, Taipei, Taiwan}
\acmPrice{15.00}




\maketitle

\section{Introduction}

Research and development in information retrieval~(IR) has been predominantly experimental. In its early days in the 1960s, the IR~community saw the need to develop and validate experimental procedures, giving rise to the Cranfield paradigm~\cite{cleverdon:1967}, which became the de facto standard for shared tasks hosted at TREC~\cite{voorhees:2019} and beyond. Organizers of typical shared IR~tasks provide a task description, a document corpus, and topics. Participants implement retrieval approaches for the task and run them on each topic to produce document rankings (a so-called ``run''). The rankings are then usually submitted as files to the organizers who pool all runs, gather (reusable) relevance judgments for the pools, and calculate the evaluation scores~\cite{voorhees:2001}. Finally, the participants describe their methodology and findings in a published ``notebook'' paper. This division of labor allowed the community to scale up collaborative laboratory experiments, especially at a time of limited bandwidths for data exchange, since run files occupy only a few kilobytes. With many research laboratories working independently on the same task, the community draws on the ``wisdom of crowds'' while ensuring rigorous comparative evaluation.

Despite the lasting success, this way of organizing shared tasks also has shortcomings. First, as with many other disciplines in computer science and beyond, the retrieval approach of a run described in a notebook paper might not be reproducible. There are well-documented cases where reproductions failed, despite putting much effort into it, even for approaches with diligently archived code repositories~\cite{arguello:2015,lin:2020}. Second, run submissions require that participants have access to the test topics, which has severe implications~\cite{fuhr:2020a}, such as informing (biasing) the research hypothesis or retrieval approach, unless researchers make a point of not looking at the topics, ever, during development. Third, it cannot be ruled out that current or future large language models have been trained, by mistake or deliberately, on publicly available test data, or that a usage warning stating not to use the data for training would go unnoticed.%
\footnote{Some form of leakage from MS MARCO~\cite{nguyen:2016} to the Flan-T5 prompting model~\cite{chung:2022} has already been observed: \href{https://web.archive.org/web/20230426101808/https://twitter.com/UnderdogGeek/status/1630983277363228672}{twitter.com/UnderdogGeek/status/1630983277363228672}, \href{https://web.archive.org/web/20230426101727/https://twitter.com/macavaney/status/1649779164625481733}{twitter.com/macavaney/status/1649779164625481733}.}
In any case, the current best practices for shared tasks do not enforce  ``blinded experimentation''%
\footnote{\href{https://en.wikipedia.org/wiki/Blinded\_experiment}{en.wikipedia.org/wiki/Blinded\_experiment}}
with sufficient rigor, compared to other empirical disciplines.

\begin{figure*}
\centering
\includegraphics[scale=.985]{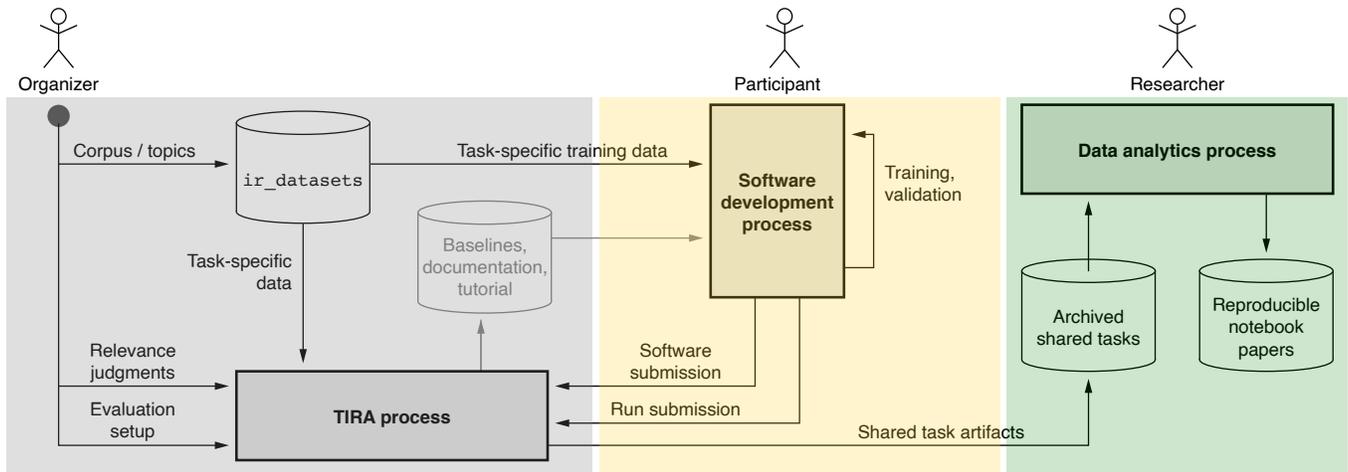}
\caption{Overview of typical shared task-like IR experiments and how the tools in TIREx support them.}
\Description{Overview of typical shared task-like IR experiments and how the tools in TIREx support them.}
\label{figure-ir-experiment-platform}
\end{figure*}

To address all of these shortcomings, we have developed the IR~Experiment Platform~(TIREx; cf.\ Figure~\ref{figure-ir-experiment-platform} for an overview). Available as open source,%
\footnote{\href{https://github.com/tira-io/ir-experiment-platform}{github.com/tira-io/ir-experiment-platform}}
a key feature of TIREx is the full integration of tools for working with IR data (\irdatasets~\cite{macavaney:2021}), for executing retrieval pipelines (PyTerrier~\cite{macdonald:2021}), and for evaluating IR systems (\irmeasures~\cite{macavaney:2022}) with the TIRA Integrated Research Architecture~\cite{froebe:2023b}, a continuous integration service for reproducible shared tasks and experiments. TIREx~is designed to for reproducibility through software submissions while keeping an experimenter's or task organizer's workload comparable to run file submissions.

On our Betaweb and Gammaweb clusters,%
\footnote{\href{https://webis.de/facilities.html\#hardware}{webis.de/facilities.html\#hardware}}
we have deployed an instance of TIREx that is open for software submissions and experiments. A substantial efficiency boost comes from integrating GPU~cores and result caching into the platform to accelerate neural IR~approaches. As a proof of concept, we conducted a large-scale evaluation of 50~``standard'' retrieval approaches on 32~shared retrieval tasks (based on 15~corpora with a total of 1.9~billion documents). This experiment consists of 1,600 runs and was started by just clicking a button. It finished unattended in less than a week.

\section{Background and Related Work}
\label{sec:related-work}

We review ad hoc retrieval experiments in evaluation campaigns, common problems and pitfalls in IR~experiments, best practices for leaderboards, existing reproducibility initiatives, and tools to support reproducibility. Insights from all these areas have influenced our implementation decisions for TIREx.

\paragraph{Ad hoc Retrieval Experiments in Evaluation Campaigns}
Today's shared task-style experiments for ad hoc retrieval evolved from the Cranfield experiments~\cite{voorhees:2019}. In the 1960s, the Cranfield experiments~\cite{cleverdon:1967,cleverdon:1991} were conducted on a corpus of 1,400~documents with complete relevance judgments for 225~topics. Since corpus sizes grew substantially, complete judgments became infeasible almost immediately thereafter~\cite{voorhees:2019}. The current practice at shared tasks in~IR thus is to only assess the relevance of per-topic pools of the submitted systems' top-ranked documents~\cite{voorhees:2019}. Subsequent evaluations on the same corpus usually are based on the assumption that the pools are ``essentially complete'', i.e., unjudged documents that were not in the pool are non-relevant~\cite{voorhees:2019}. Although this completeness assumption is reasonable for tasks with a diverse set of submitted runs pooled at high depth~\cite{voorhees:2021}, recent observations suggest that scenarios with many relevant documents per query (e.g., corpora with many duplicates~\cite{voorhees:2022}) or with topics representing broad information needs~\cite{thakur:2021} are rather problematic. Especially for shared tasks that do not attract diverse submissions, TIREx can help to produce a more diverse judgment pool, as a wide range of different baseline retrieval systems is directly available and can be applied to any imported retrieval task.

\paragraph{Common Problems and Pitfalls in IR~Experiments}
Even though the current discussion about how to conduct IR~experiments~\cite{fuhr:2017,sakai:2020,zobel:2023} includes some controversial points (e.g., whether MRR should be abandoned~\cite{fuhr:2017} or not~\cite{sakai:2020,moffat:2022}), there is still a consensus in the IR~community on many characteristics of ``bad'' or ``good'' experiments. For instance, it is rather undisputed that retrieval studies should be internally valid (conclusions must be supported by the data) and externally valid (repeating an experiment on different but similar data should yield similar observations)~\cite{fuhr:2020b}. Still, external validity of IR~experiments remains an open problem~\cite{fuhr:2020a}. TIREx~can help to further improve both: the internal validity via archiving all experiments and results on some corpus (e.g., to accurately correct for multiple hypothesis tests), and the external validity via simplifying to run a submitted software on different data. 

\citet{thakur:2021} attempted to address the external validity problem by combining diverse retrieval corpora in the BEIR~benchmark for \emph{en masse} evaluation. However, in practice, running an approach on all corpora in~BEIR requires some effort, so that many studies still only report results for a selected subset (e.g.,~\cite{gao:2022,formal:2021,bonifacio:2022})---often even without clearly justifying the selection. In contrast, a software in TIREx can rather easily be evaluated against many if not all corpora so that analyzing improvements and limitations of an approach on diverse data is not much effort.

An often criticized practice is that many IR~studies compare a new approach against weak or ``wrong'' baselines (i.e., not the best or most reasonable previous approaches). Any improvements claimed in such studies are not really meaningful~\cite{armstrong:2009b,lin:2018}. One reason for choosing a wrong baseline could be that neither the researchers nor the reviewers are actually aware of what previous approaches exist for a specific corpus since results are often scattered across multiple publications~\cite{lin:2018}. Centralized leaderboards that directly show the effectiveness of diverse approaches for a wide range of tasks would address this problem, but multiple efforts have failed so far~\cite{lin:2018}. In~TIREx, we include many popular corpora and standard retrieval approaches right from the start so that the TIREx~leaderboards can initially gain traction. The more shared tasks (but also researchers) employ TIREx for software submissions, the broader TIREx' coverage will get over time.

\paragraph{Maintaining Ongoing Leaderboards}
Inspired by the observation that many IR~studies do not compare a new approach against reasonable baselines (e.g., the most effective TREC~runs)~\cite{armstrong:2009b}, \citet{armstrong:2009} released EvaluateIR, a public leaderboard accepting run file submissions. Although the concept was highly valuable for the community in helping researchers and reviewers alike to select appropriate baselines, ``EvaluateIR never gained traction, and a number of similar efforts following it have also floundered''~\cite{lin:2018}.

While there is still no centralized general leaderboard for~IR, certain task-specific leaderboards are quite popular. For instance, the leaderboard of the recent MIRACL~Challenge~\cite{zhang:2022} received 25~submissions within one week, and the MS~MARCO leaderboard~\cite{lin:2022b} has been popular for years. Maintaining such long-running leaderboards comes with some caveats, as they are conceptually turn-based games where every leaderboard submission might leak information from the test set~\cite{lin:2022b}. \citet{lin:2022b} propose best practices, inspired by previous problems of the Netflix prize.%
\footnote{\href{https://web.archive.org/web/20110225015552/http://www.netflixprize.com/}{www.netflixprize.com}} %
Most importantly, \citeauthor{lin:2022b} note that, while submissions to a leaderboard are open, the retrieval results should not be public, nor should system descriptions or implementations, as this would potentially leak information from the test set and foster ``uninteresting'' approaches like ensembles of all the top submissions. With TIREx and its blind evaluation, organizers can choose to blind all submissions as long as they need to, with the ability to unblind approaches and submissions as they see fit, so that TIREx supports the best practices recommended by \citet{lin:2022b}.

\paragraph{Reproducibility Initiatives in IR}
Reproducibility is a major challenge in research. For instance, a survey among 1,576~researchers revealed that more than 50\% failed at least once to reproduce their own experiments~\cite{baker:2016}. The IR community makes substantial efforts to foster reproducibility. There are, for instance, dedicated reproducibility tracks at conferences%
\footnote{Examples at ECIR~2023 and SIGIR~2023: \href{https://web.archive.org/web/20230217074212/https://ecir2023.org/calls/reproducibility.html?v=1.22}{ecir2023.org/calls/reproducibility.html} and \href{https://web.archive.org/web/20221220020314/https://sigir.org/sigir2023/submit/call-for-reproducibility-track-papers/}{sigir.org/sigir2023/submit/call-for-reproducibility-track-papers}.} %
and dedicated reproducibility initiatives like OSIRRC~\cite{arguello:2015,clancy:2019} or CENTRE~\cite{ferro:2018,ferro:2019,sakai:2019,sakai:2020b}. OSIRRC aims to produce archived versions of retrieval systems that are replicable, while CENTRE runs replicability and reproducibility challenges across IR~evaluation campaigns. \citet{lin:2020} looked at all the artifacts produced in the OSIRRC~2015 challenge~\cite{arguello:2015} to verify which results are still replicable four years after their creation. Out of the seven systems that participated in the challenge, only the results of Terrier~\cite{ounis:2005} were fully reproducible out of the box, while two other systems could still be fixed by manual adjustments to the code. The main reasons for failure were that external dependencies could not be loaded anymore, or that platform dependencies changed (i.e., the operating system with its packages). To mitigate the problem of changing platform dependencies, the follow-up iteration of OSIRRC~\cite{clancy:2019} focused on Docker images that had to implement a strict specification (enforced by the companion tool ``jig'') that triggered the indexing and subsequent retrieval via Docker hooks. Even though 17~systems have been dockerized to follow the jig specification, the concept has not gained traction. By centering TIREx around shared tasks in the beginning, we hope that we can kick off and maintain the attention of the community. Furthermore, we believe that there are many retrieval scenarios that can not be encapsulated into the two-step index-then-retrieve pipeline that jig imposes (e.g., explicit relevance feedback). We thus minimize the TIREx~requirements: just Docker images in which commands are executed without Internet access on read-only mounted data.

\paragraph{Tooling for Reproducibility}
Many tools have been developed to support shared tasks by reducing the workload of organizers and participants while increasing the reproducibility~\cite{yadav:2021,breuer:2019,vanschoren:2013,jagerman:2018,tsatsaronis:2015,hopfgartner:2015,froebe:2023b}. For instance, as documenting the metadata of experiments improves reproducibility~\cite{leipzig:2021}, \irmetadata~\cite{breuer:2022} simplifies the documentation of IR~experiments according to the PRIMAD~model~\cite{ferro:2016} (platform, research goal, implementation, method, actor, data). There are also platforms that support organizing and running shared tasks, among which four are still active: CodaLab, EvalAI, STELLA, and TIRA.%
\footnote{\raggedright\href{https://codalab.org}{codalab.org}, \quad \href{https://eval.ai}{eval.ai}, \quad \href{https://stella-project.org}{stella-project.org}, \quad \href{https://tira.io}{tira.io}}
They implement the so-called evaluation-as-a-service paradigm in the form of cloud-based web services for evaluations~\cite{hopfgartner:2018}. Of these four systems, STELLA and TIRA are hosted within universities, while CodaLab and EvalAI use Microsoft Azure and Amazon~S3, respectively. We use TIRA for TIREx as it supports blinded experimentation and as it is based on (private) git repositories hosted on GitLab or GitHub to versionize shared tasks and to distribute the workloads via runners connected to the corresponding repositories. The computation can thus be done in the cloud but also on private machines. We substantially extend large parts of TIRA as part of TIREx so that it supports the current IR~workflows like chaining multiple retrieval stages.

\section{The IR Experiment Platform}
\label{sec:ir-experiment-platform}

We have constructed the Information Retrieval Experiment Platform (TIREx) to facilitate reproducible, shared task-style IR~experiments based on software submissions. This has been achieved by integrating \irdatasets, \irmeasures, and PyTerrier into TIRA. We anticipate the sustained availability and maintenance of these components, as evidenced by TIRA's and PyTerrier's consistent upkeep since 2012~\cite{gollub:2012c} and 2020~\cite{macdonald:2020}, respectively, and the growing popularity of \irdatasets in recent years. Previously, conducting shared task-style IR~experiments within TIRA was already possible, but required significant effort from both organizers and participants due to their unique nature, compared to standard Machine Learning or Natural Language Processing experiments. IR~experiments typically involve intermediate artifacts (like indexes), and retrieval systems involve multi-stage ``telescoping'' pipelines.%
\footnote{For instance, the Mono-Duo-Reranking pipelines~\cite{pradeep:2021}, where a more complex re-ranker improves part of the ranking of a less complex one ahead in the pipeline.}
To address these requirements, TIREx extends TIRA with common IR~tools for data access, indexing, retrieval, and evaluation, and implements multi-stage pipelines on top of TIRA's underlying execution protocol. Below, we elaborate on how TIREx supports IR~experiments, discuss the interaction between integrated tools, provide examples of using available retrieval approaches in TIREx, and demonstrate how TIREx promotes post-experiment replicability and reproducibility through declarative PyTerrier pipelines.

\subsection{Experiments in the IR Experiment Platform}

As illustrated in Figure~\ref{figure-ir-experiment-platform}, TIREx facilitates the entire process of conducting retrieval experiments. It allows shared task organizers and individual experimenters to import data and utilize any pre-existing retrieval software submitted to TIREx as baselines. Following that, submissions of new retrieval approaches for evaluation can be made as software submissions or, if enabled, also as run submissions. Any submission can be accompanied by descriptive annotations and metadata; for instance, run submissions can be grouped to denote that they were generated by the same retrieval approach for multiple retrieval tasks. By providing relevance judgments, organizers or experimenters can directly evaluate all available runs.

To incorporate a new corpus and topics into TIREx, they can be easily added to \irdatasets, utilizing a private branch if the data is sensitive. This data can then be imported by TIRA through a Docker image with a matching \irdatasets installation. Participants submit their software as Docker images as well. TIRA ensures their reproducibility and prevents test data leaks by executing them in a sandbox. Among other things, the sandbox disables Internet connectivity for the running software, which ensures that the software and its dependencies are fully installed and no data is sent to unauthorized third parties. Participants can provide additional data their software needs during execution by uploading it to TIRA. This is particularly useful for non-reproducible elements of a submission, such as manual query reformulations. TIREx also provides a ``starter implementation'' for five commonly used IR~research frameworks, which participants can use as a development base. The simplest starter uses BM25~retrieval, which is implemented using a few lines of declarative PyTerrier code in a Jupyter notebook.%
\footnote{\href{https://github.com/tira-io/ir-experiment-platform\#starter-for-pyterrier-in-jupyter}{github.com/tira-io/ir-experiment-platform\#starter-for-pyterrier-in-jupyter}}

TIREx allows for software submissions to be executed on demand within a cloud-based execution environment, utilizing GitLab or GitHub CI/CD pipelines. In order to meet varying demand, experiment organizers can incorporate additional runners as necessary. TIREx maintains a comprehensive record of every artifact of a retrieval experiment within a specific git repository (Figure~\ref{figure-ir-experiment-platform}, right), which can be exported and published. This ``archived shared task'' is entirely self-contained, enabling the independent re-execution of approaches with identical or differing data using PyTerrier pipelines. The availability of every software that generated a run as part of the repository makes it a key outcome and asset of an experiment. Consequently, TIREx facilitates ``always-on'' shared tasks for the IR~community, along with an extensive variety of ablation studies.

\subsection{Reproducible Shared Tasks with TIRA}

TIRA is used to handle software submissions in shared tasks since~2012~\cite{gollub:2012c,potthast:2019p}---the CLEF~labs PAN and Touch{\'e}
being two long-running examples.%
\footnote{\href{https://pan.webis.de/}{pan.webis.de}, \quad \href{https://touche.webis.de/}{touche.webis.de}}
A first version of TIRA provided participants with access to virtual machines to deploy their software. However, this setup required manual overhead on the part of organizers thus did not scale far beyond these two events. Moreover, software reexecution was possible in principle and has been demonstrated once at scale~\cite{hagen:2017c}, but proved to be error-prone and required manual bug fixing inside the virtual machines as participant software was not robust against slight data format variations that were in principle supported by the underlying formatting schema. This also has prevented external researchers from reproducing the collected software for a given task at scale.

Meanwhile, Docker has gained maturity and widespread adoption and is now supported by many cluster computing frameworks such as Kubernetes. Especially their integration as GitHub and GitLab runners made automatic deployment widely available. Hence, TIRA was completely redeveloped based on the now industry-standard CI/CD~pipelines (continuous integration and deployment) using Git, Docker, and Kubernetes~\cite{froebe:2023b}. In the new version of TIRA, participants upload their software implemented in Docker images to a private Docker registry dedicated to their team, ensuring that different teams do not influence each other while a shared task is running---the approaches can remain private until the task ends. For on-demand execution, TIRA presently runs the software on our Kubernetes cluster (1,620~CPU cores, 25.4\,TB RAM, 24~GeForce GTX 1080~GPUs). This version of TIRA was first used in two NLP~tasks hosted at SemEval~2023 to which 71~of 170~registered teams submitted 647~runs based on software submissions~\cite{froebe:2023d,kiesel:2023a}. 

While preparing the TIRA setup for the retrieval-oriented Touch{\'e}~2023 tasks~\cite{bondarenko:2023a}, we realized that the new TIRA still had some shortcomings. There was no unified access to IR~data, no separation between full-rank or re-rank approaches, no modularization of software components with caching, and typical IR~workflows were only realizable inefficiently or via workarounds. For instance, full-rank retrieval in TIRA would have required any software to build an index from scratch and different re-rank approaches would each have to re-create the baseline rankings. A re-ranking approach for the ClueWeb22-based Task~2 of Touch{\'e}~2023~\cite{bondarenko:2023a}, for example, should have been able to use a ChatNoir baseline ranker~\cite{bevendorff:2018} from within TIRA, but our pilot experiments showed that retrieving the top-1000~ChatNoir results for some set of 50~Touch{\'e} topics~\cite{bondarenko:2020e,bondarenko:2021d,bondarenko:2022f,bondarenko:2023a} takes 54~to 134~minutes (ChatNoir requests can fail so that a client has to retry the requests). Blocking GPUs---often required by re-rankers---for such a long time would waste resources and the baseline's top-1000 results should ideally be cached so that different re-rankers can directly use them. To solve all these problems, we substantially expanded TIRA and redeveloped major parts to integrate \irdatasets, \irmeasures, and PyTerrier.

\subsection{Standardized Data Access with \texttt{ir\_datasets}}

The \irdatasets toolkit~\cite{macavaney:2021} provides a standard interface to access over 200~corpora and over 500~topic sets frequently used in IR~experiments. The data is kept up-to-date (e.g., most TREC~2022 tracks are included) and processing documents or topics is possible via a single line of Python code. Thus, \irdatasets already serves as a common data layer in numerous IR~frameworks and tools (e.g., Capreolus~\cite{yates:2020}, Experimaestro-IR~\cite{piwowarski:2020}, FlexNeuART~\cite{boytsov:2020}, OpenNIR~\cite{macavaney:2020}, Patapsco~\cite{costello:2022}, PyTerrier~\cite{macdonald:2021}) and can be easily incorporated by most others (e.g., Anserini~\cite{yang:2017}, PISA~\cite{mallia:2019}). We integrate \irdatasets into TIRA via Docker images that can import complete corpora (for full-rank approaches) and that can create re-rankings for any given run file (for re-ranking approaches). To configure an IR~experiment in TIRA, the experiment organizer only needs to provide an \irdatasets Docker image---standard images are available in~TIREx but other images are also possible (e.g., for proprietary data). In the following, we further describe the new `default\_text' fields that we added to \irdatasets to enable re-using single-field retrieval software on corpora with multiple text fields, and we describe how the integration of \irdatasets into Docker images that run on-demand also ensures interchangeability and compatibility of retrieval components in retrieval pipelines.

\begin{table*}
\centering
\small
\renewcommand{\arraystretch}{1}%
\setlength{\tabcolsep}{7.75pt}%
\caption{Overview of what data TIRA makes available to full-rank and re-rank approaches. The `Access' columns indicate the default accessibility to participants~(P), organizers (O; can make data accessible as indicated by $\dagger$), and unregistered users~(U).}
\label{table-data-interfaces}
\vspace*{-2ex}
\begin{tabular}{@{}lllcccl@{}}
\toprule
\bfseries Type             & \bfseries Resource                  & \bfseries Fields                                 & \multicolumn{3}{@{}c@{}}{ \bfseries Access}                                             & \bfseries Example Entry \\
\cmidrule(l@{\tabcolsep}r@{\tabcolsep}){4-6}
                           &                                     &                                                  &            P            &            O            &                  U                  & \\
\midrule
\multirow{3}{*}{Full-Rank} & \multirow{2}{*}{documents.jsonl.gz} & \multirow{2}{*}{docno, text, original\_document} & \multirow{2}{*}{\cmark} & \multirow{2}{*}{\cmark} & \multirow{2}{*}{\xmark$^{\dagger}$} & \multirow{2}{*}{\footnotesize \shortstack[l]{\{"docno": "8182161", "text": "Goldfish can grow up to 18 inches \ldots",\\\phantom{\{}"original\_document": \{\ldots\}\}}} \\
                           &                                     &                                                  &                         &                         &                                     & \\
\cmidrule(l@{\tabcolsep}){2-7}
                           & topics.jsonl.gz                     & qid, query, original\_topic                      &         \cmark          &         \cmark          &         \xmark$^{\dagger}$          & {\footnotesize \{"qid": "156493", "query": "do goldfish grow", "original\_query": \{\ldots\}\}} \\
\midrule
\multirow{3}{*}{Re-Rank}   & \multirow{3}{*}{re-rank.jsonl.gz}   & qid, query, original\_topic, docno,              &                         &                         &                                     & \multirow{3}{*}{\footnotesize \shortstack[l]{\{"qid": "156493", "query": "do goldfish grow", "original\_query": \{\ldots\}, \\ \phantom{\{}"docno": 8182161, "text": "Goldfish can grow up to 18 inches \ldots",\\  \phantom{\{}"original\_document": \{\ldots\}, "rank": 1, "score": 31.16\}}} \\
                           &                                     & text, original\_document                         &         \cmark          &         \cmark          &         \xmark$^{\dagger}$          & \\
                           &                                     & score, rank                                      &                         &                         &                                     & \\
\midrule
Both                       & qrels.txt                           & topic, iteration, docno, relevance               &   \xmark$^{\dagger}$    &         \cmark          &         \xmark$^{\dagger}$          & {\footnotesize 156493 Q0 8182161 2} \\
\bottomrule
\end{tabular}
\end{table*}

\paragraph{Re-Usable Retrieval Software via default\_text}
While some corpora have a single text field for each document (e.g., the MS~MARCO passage ranking corpus~\cite{nguyen:2016,craswell:2019,craswell:2020}), others provide rich structural information or metadata (e.g., the Touch{\'e} corpora~\cite{bondarenko:2021d,bondarenko:2022f} with structured arguments or comparison aspects). Similarly, some retrieval tasks have a single text field per topic (e.g., Antique~\cite{hashemi:2020}), while others provide metadata for each topic and/or multiple fields for versions of a query (e.g., TREC Precision Medicine~\cite{roberts:2017,roberts:2018}).

Corpora and retrieval tasks with fine-grained structure usually address the development of built-for-purpose retrieval systems that exploit the task-specific setup. For instance, an argument retrieval system submitted to Touch{\'e} may specifically focus on the argumentative premises contained in a document, and an approach in the Precision Medicine track may use a query's structure to adjust the relevance criteria. Instead, corpora and tasks with single fields for document texts and queries often rather address ``general search'' scenarios (i.e., retrieval approaches that can be applied in a variety of contexts rather than targeting one specific case). To also enable the evaluation of such general purpose retrieval systems (that expect a single document text field and a single query field) on data with more fields, we created default\_text fields for every dataset in \irdatasets. There often is a natural choice for a document's or a query's ``default text'' (e.g., we simply concatenated the two fields `title' and `abstract' of MEDLINE~documents as the default document text and we often selected a TREC topic's title as the query text---after a manual review). Still, there also are more difficult cases for which we then carefully tried to select the most important content of the documents or topics---being open to corrective pull requests from the community. The new default\_text fields now are part of the \irdatasets package and thus also applicable in~TIREx to ensure reusability of single-field retrieval approaches on data originally only available with multiple fields.

\paragraph{Ensuring Compatibility of Modularized Retrieval Stages}
TIREx aims to support experiments in which components for the individual stages of modularized retrieval pipelines can be easily replaced and compared without having to adapt the complete retrieval software each time. Therefore, TIRA distinguishes between two types of retrieval approaches:
\Ni
full-rank approaches with a document corpus and topics as input, and
\Nii
re-rankers with a re-rank file as input (basically, query--document pairs).
From any retrieval software's output, a re-rank file can be automatically created and cached in TIREx by the \irdatasets integration. As the structure of these re-rank files always is the same, any re-ranker can easily run on the output of any previous retrieval approach. Note that some data in \irdatasets can not be downloaded from the Web and/or requires license agreements (e.g., the ClueWeb and GOV corpora). As we have valid license agreements on our local TIREx~instance, we can directly mount such data into the \irdatasets container, but, by default, then only show effectiveness scores for a run and no retrieval results (i.e., participants do not get access to the corpus as their software is executed in a sandbox and all outputs other than effectiveness scores are not shown on confidential datasets).

Table~\ref{table-data-interfaces} shows the data fields that the \irdatasets integration makes available. For full-rank software, the \textttsmall{documents.jsonl.gz} file for each document contains an identifier `docno', the new default\_text in the field `text', and all original structured fields of a document in `original\_document'. The \textttsmall{topics.jsonl.gz} file for each topic contains an identifier `qid', the new default\_text in the field `query', and all original structured fields of a topic in `original\_topic'. For re-rankers, the \irdatasets integration creates a file \textttsmall{re-rank.jsonl.gz} from the output of a previous retrieval stage (i.e., the run file), where each entry contains query--document pairs to be reranked along their score and rank assigned by the previous stage. When relevance judgments exist, the \irdatasets integration can also make them available in a \textttsmall{qrels.txt} file so that the evaluator software specified by the experiment organizer can automatically evaluate submitted retrieval approaches.

\subsection{Sanity-checked Evaluation with \texttt{\fontsize{10pt}{11pt}\selectfont ir\_measures}}

TIRA can automatically evaluate run files (created by retrieval software submissions or uploads) via an \irmeasures evaluator. First, the evaluator performs a sanity check to test whether a run file can be parsed and warns of potential errors (e.g., score ties, NaN~scores, empty result sets, unknown queries, scores contradicting the ranks, etc.). Then, if relevance judgments have been provided, the evaluator derives all specified measures averaged over all queries and per query (suitable for significance tests).

\subsection{Reproducible IR Pipelines with TIRA}

To improve the efficiency of common IR~workflows in~TIREx, we redeveloped and extended TIRA's ability to define and run modularized software even spanning multiple Docker images. All software in TIRA is immutable so that outputs of one software (e.g., an index) can be cached and reused by another software.

\begin{table}
\centering
\small
\renewcommand{\arraystretch}{1}%
\setlength{\tabcolsep}{6.5pt}%
\caption{Overview of variables available for software in TIRA. The {\textttsmall{\$inputDataset}\xspace} and {\textttsmall{\$outputDir}\xspace} variables are always available, while {\textttsmall{\$inputRun}\xspace} is only available for multi-component software depending on previous stages.}
\label{table-data-injection-in-tira}
\vspace*{-2ex}
\begin{tabular}{@{}lll@{}}
\toprule
\bfseries Variable                    & \bfseries Availability & \bfseries Description                \\
\midrule
{\textttsmall{\$inputDataset}\xspace} & Always                 & Directory containing the input data. \\
{\textttsmall{\$outputDir}\xspace}    & Always                 & Directory with expected output data. \\
{\textttsmall{\$inputRun}\xspace}     & Multi-Comp.            & Output(s) of previous stage(s).      \\
\bottomrule
\end{tabular}
\end{table}

\paragraph{Modularized Software with Multiple Components}
Retrieval software in TIRA can have multiple components that form a sequence similar to UNIX~pipes or even a directed acyclic graph~(DAG). Each component has a Docker image with a command to be executed and can have none, one, or many preceding components, respectively. TIRA passes the corresponding input and output directories to each component via three variables (cf.\ Table~\ref{table-data-injection-in-tira}). The variable \textttsmall{\$inputDataset} points to the directory that contains the actual input (e.g., \textttsmall{re-rank.jsonl.gz} for re-ranking software). The variable \textttsmall{\$inputRun} is only available when a component has preceding components and then points to a directory with all outputs of the directly preceding components. The variable \textttsmall{\$outputDir} specifies the location where TIRA expects a component's outputs. All three variables \textttsmall{\$inputDataset}, \textttsmall{\$inputRun}, and \textttsmall{\$outputDir} can be included in the command to be executed but are also available as environment variables within a container.

\paragraph{Components for Additional Data}
A retrieval approach might use data unavailable in an experiment's original corpus and topics. An example are user query variants for ClueWeb or Common Core topics~\cite{bailey:2016,benham:2018}---using the variants would count as a ``manual'' run at~TREC. TIREx supports such cases of additional data via file uploads which becomes available to subsequent software via the \textttsmall{\$inputRun} variable. Uploaded files can be grouped, documented, and configured as components treated like a software component (e.g., to precede some ranking component). Thus, TIREx supports manual runs and other kinds of use cases, but isolates such steps as much as possible to keep the software part of a pipeline replicable.

\begin{figure}[tb]
\centering
\includegraphics[width=\columnwidth]{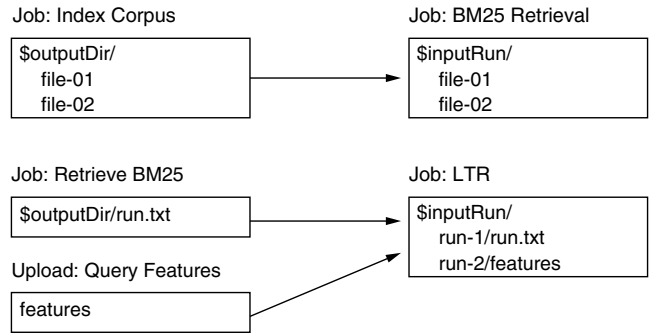}
\caption{Data flow of two retrieval pipelines in TIRA. The upper retrieval pipeline creates an index so that the second stage retrieves from the index with BM25. The bottom retrieval pipeline uses a BM25 ranking and a manually uploaded file with query features as input for an LTR algorithm.}
\Description{Data flow of two retrieval pipelines in TIRA. The upper retrieval pipeline creates an index so that the second stage retrieves from the index with BM25. The bottom retrieval pipeline uses a BM25 ranking and a manually uploaded file with query features as input for an LTR algorithm.}
\label{figure-multi-stage-pipelines}
\end{figure}

\paragraph{Defining Retrieval Pipelines}
Figure~\ref{figure-multi-stage-pipelines} illustrates the conceptual data flow for the two simplest-possible sequence and DAG retrieval pipelines, respectively. The upper pipeline shows a full-rank approach that first creates an index (component `Index Corpus' with output \textttsmall{file-01} and \textttsmall{file-02}) that TIRA then makes available as \textttsmall{\$inputRun} for the second component `BM25 Retrieval'. Figure~\ref{tira-define-job} shows how a BM25~retrieval component that depends on a PyTerrier index can be defined in~TIRA. Since many different components of a software may use a created artifact like an index, we cache all outputs to make pipelines more efficient. 

\bsfigure[width=\columnwidth]{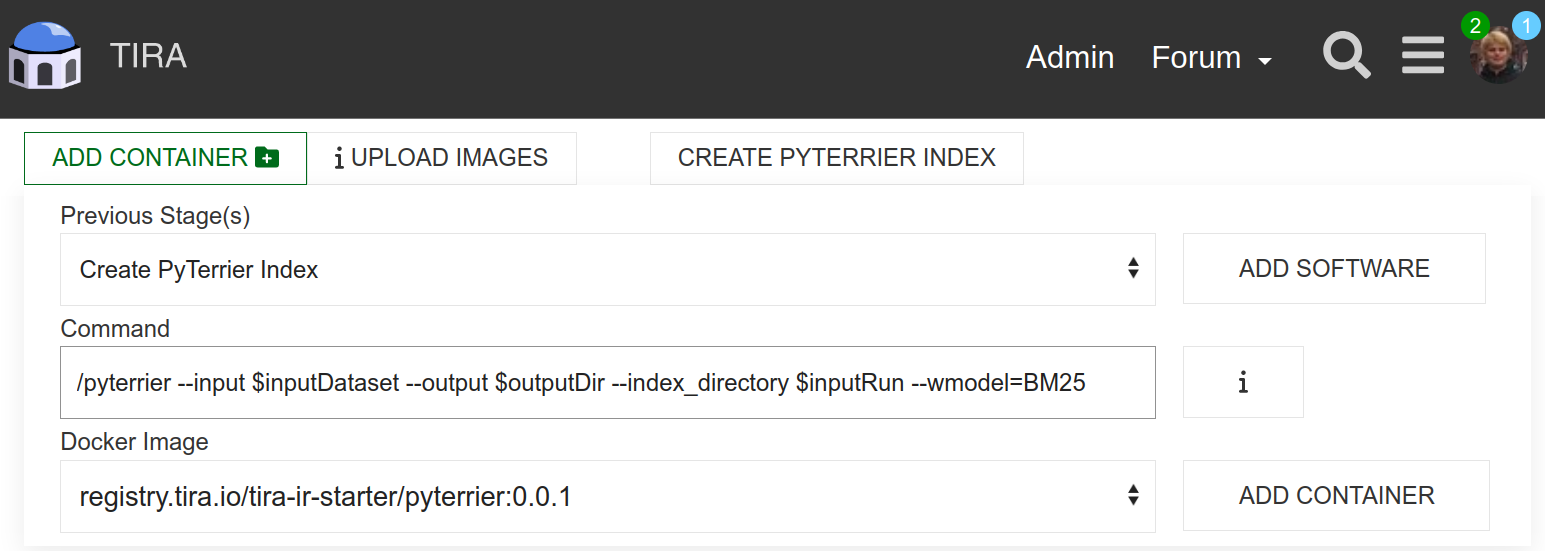}{Defining a BM25~retrieval component in TIRA that depends on a previously created PyTerrier index.}

The lower pipeline in Figure~\ref{figure-multi-stage-pipelines} shows a learning-to-rank component that depends on a BM25~retrieval and on uploaded query features (e.g., user behavior data like clicks and dwell-times obtained from a user study). When a component has inputs from more than one component, TIRA makes them available in the order in which the components have been defined. Preceding components or uploads must exist when defining a new component and TIRA~decouples the command to be executed from the Docker image so that the same image can be used to run different retrieval approaches (e.g., by switching parameters). In combination with caching, this improves the efficiency for a wide range of common multi-stage retrieval pipelines.

\paragraph{Efficiency via Caching}
As every software in TIRA is immutable, coding errors in a component can only be fixed by adding a new version of the component. Immutability enables the implementation of efficient and reliable retrieval pipelines, since their output can both be cached and traced back and replicated by that same (version) of a component. TIRA disallows the deletion of components or outputs that have been used as inputs by some other component. When any component is requested to produce an output on some data, it is first checked whether that output already is cached in which case executing the component is not necessary. This way, retrieval pipelines in TIRA can efficiently re-use components and remain replicable, as the steps to produce a final run are fully tracked and versioned in the experiment repository.

\subsection{Local Pipeline Reproduction with PyTerrier}
\label{sec:reproducibility-notebooks}

When an experiment repository is exported and published by the organizers, by default, the test data is kept private but the run files are published via TIRA and software submissions are uploaded as Docker images to Docker~Hub. All possible follow-up studies (e.g., a reproducibility study for a shared task) can be conducted independent of TIRA, as archived experiment repositories are fully self-contained. In the following, we briefly showcase some post-hoc experiments in PyTerrier.%
\footnote{Examples available at: \href{https://github.com/tira-io/ir-experiment-platform\#reproducibility}{github.com/tira-io/ir-experiment-platform\#reproducibility}}

\begin{listing}[tb]
\begin{lstlisting}[firstline=2, caption=Full-rank retrieval from a complete corpus., label=listing-corpus-retrieval]
pipeline = tira.pt.retriever(
    '<task-name>/<user-name>/<software>',
    dataset='<dataset>'
)
advanced_pipeline = pipeline >> advanced_reranker
\end{lstlisting}
\end{listing}

Listing~\ref{listing-corpus-retrieval} shows how a full-rank approach from a TIRA~experiment repository can be reproduced with a declarative PyTerrier pipeline. The approach is identified as the \textttsmall{<software>} submitted by team \textttsmall{<user-name>} to the shared task \textttsmall{<task-name>} and is applied to \textttsmall{<dataset>} (does not need to be the original task data). Internally, the required Docker images are downloaded and run in their required order to obtain the results. These results can then be re-ranked by any PyTerrier re-ranker, allowing for experiments to improve an original submission. Also re-rankers available in some TIRA~experiment repository can be used in post-hoc PyTerrier experiments (cf.\ Listing~\ref{listing-rerank} for an example re-ranking of~BM25).

\begin{listing}[tb]
\begin{lstlisting}[firstline=2, caption= Re-ranking BM25 with a submitted software., label=listing-rerank]
bm25 = pt.BatchRetrieve(index, wmodel="BM25")
reranker = bm25 >> tira.pt.reranker(
    '<task-name>/<user-name>/<software>'
)
\end{lstlisting}
\end{listing}

Listing~\ref{listing-pyterrier-reranker} shows how run files resulting from some (software) submission can be loaded into PyTerrier. The  \textttsmall{from\_submission} method allows to access some submitted approach's output without having to re-run it (e.g., this also eases pooling for task organizers). The PyTerrier integration allows easy replicability experiments if the dataset is the same as in the original experiment, and reproducibility experiments if some other dataset is used for retrieval approaches.

\begin{listing}[tb]
\begin{lstlisting}[firstline=2,caption=Re-ranking a run created by a software submission.,label=listing-pyterrier-reranker]
first_stage = tira.pt.from_submission(
    '<task-name>/<user-name>/<software>',
    dataset='<dataset>'
)
advanced_pipeline = first_stage >> advanced_reranker
\end{lstlisting}
\end{listing}

\section{Evaluation}
\label{sec:evaluation}

To demonstrate the scalability of TIREx, we report about an experiment with 50~retrieval approaches on 32~retrieval tasks based on 15~corpora (1.9~billion documents). The resulting leaderboards are public and new submissions can be made at any time.%
\footnote{\href{https://github.com/tira-io/ir-experiment-platform\#submission}{github.com/tira-io/ir-experiment-platform\#submission}}
We also describe a \reproeval-based~\cite{breuer:2021} case study on system preference reproducibility for different tasks.

\subsection{Scalable Retrieval Experiments}

\begin{table}
\centering
\small
\renewcommand{\arraystretch}{1}%
\setlength{\tabcolsep}{2.5pt}%
\caption{The 15~corpora and the associated 32~retrieval tasks currently available in TIREx (all are open for submissions).}
\label{table-overview-tasks}
\vspace*{-2ex}
\begin{tabular}{@{}l@{}rrl@{}r@{}}
\toprule
      \multicolumn{3}{c@{}}{\bfseries Corpus}       & \multicolumn{2}{@{}c@{}}{\bfseries Associated Retrieval Tasks }                                         \\
\cmidrule(r@{\tabcolsep}){1-3}\cmidrule(l@{\tabcolsep}){4-5}
Name                & Docs.\phantom{\,} &      Size & Details                                                                                            & \# \\
\midrule
Args.me             &            0.4\,m &   8.3\,GB & Touch{\'e}~2020--2021~\cite{bondarenko:2020e,bondarenko:2021d}                                     &  2 \\
Antique             &            0.4\,m &  90.0\,MB & QA Benchmark~\cite{hashemi:2020}                                                                   &  1 \\
ClueWeb09           &            1.0\,b &   4.0\,TB & Web tracks 2009--2012~\cite{clarke:2009,clarke:2010,clarke:2011,clarke:2012}                       &  4 \\
ClueWeb12           &          731.7\,m &   4.5\,TB & Web tracks~\cite{thompson:2013,thompson:2014}, Touch{\'e}~\cite{bondarenko:2021d,bondarenko:2022f} &  4 \\
ClueWeb22B          &          200.0\,m &   6.8\,TB & Touch{\'e}~2023~\cite{bondarenko:2023a} (ongoing)                                                  &  1 \\
CORD-19             &            0.2\,m &   7.1\,GB & TREC-COVID~\cite{voorhees:2020,wang:2020}                                                          &  1 \\
Cranfield           &             1,400 &   0.5\,MB & Fully Judged Corpus~\cite{cleverdon:1967,cleverdon:1991}                                           &  1 \\
Disks4+5            &            0.5\,m & 602.5\,GB & TREC-7/8~\cite{voorhees:1998,voorhees:1999}, Robust04~\cite{voorhees:1996,voorhees:2004}           &  3 \\
GOV                 &            1.2\,m &   4.6\,GB & Web tracks 2002--2004~\cite{craswell:2002,craswell:2003,craswell:2004}                             &  3 \\
GOV2                &           25.2\,m &  87.1\,GB & TREC TB 2004--2006~\cite{clarke:2004,clarke:2005,buettcher:2006}                                   &  3 \\
MEDLINE             &            3.7\,m &   5.1\,GB & TREC Genomics~\cite{hersh:2004,hersh:2005}, PM~\cite{roberts:2017,roberts:2018}                    &  4 \\
MS~MARCO            &            8.8\,m &   2.9\,GB & Deep Learning 2019--2020~\cite{craswell:2019,craswell:2020}                                        &  2 \\
NFCorpus            &             3,633 &  30.0\,MB & Medical LTR Benchmark~\cite{boteva:2016}                                                           &  1 \\
Vaswani             &            11,429 &   2.1\,MB & Scientific Abstracts                                                                               &  1 \\
WaPo                &            0.6\,m &   1.6\,GB & TREC Core 2018                                                                                     &  1 \\
\midrule
$\sum$ = 15 corpora &            1.9\,b &  15.3\,TB &                                                                                                    & 32 \\
\bottomrule
\end{tabular}
\end{table}

Table~\ref{table-overview-tasks} shows the 15~corpora currently available in TIREx. Each has been used for~1 to~4 shared retrieval tasks, consists of~1,400 to 1~billion documents, and comes with the relevance judgments created during the respective shared tasks.

Table~\ref{table-overview-retrieval-software} overviews the 50~retrieval approaches that we imported into TIREx from 5~retrieval frameworks: BEIR~\cite{thakur:2021}, ChatNoir~\cite{bevendorff:2018}, Pyserini~\cite{lin:2021} (our import was not ready during the experiments), PyGaggle~\cite{lin:2021}, PyTerrier~\cite{macdonald:2021} (including two PyTerrier plugins for duoT5~\cite{pradeep:2021} and ColBERT~\cite{khattab:2020}). From BEIR, we use 17~dense retrieval approaches (e.g., ANCE~\cite{xiong:2021}, DPR~\cite{karpukhin:2020}, and TAS-B~\cite{hofstatter:2021}) by using the different SBERT~\cite{reimers:2019} models available in BEIR. ChatNoir is an Elasticsearch-based BM25F search engine hosting all three ClueWeb corpora. It can be accessed from within TIRA to allow retrieval approaches on huge corpora with a REST-API that is kept consistent to ensure reproducibility. From Pyserini, we use the 4~lexical models available trough the SimpleSearcher interface. From PyGaggle, we use 8~variants of monoBERT~\cite{frassetto:2019} and monoT5~\cite{pradeep:2021} (including the state-of-the-art monoT5 with 3~billion parameters), and from PyTerrier, we use 20~lexical retrieval models (e.g., BM25, PL2, etc.). From the duoT5 plugin of PyTerrier, we use 3~variants based on different duoT5 models (including the state-of-the-art model with 3~billion parameters). For all retrieval approaches, we keep all parameters at their default values. Almost all approaches use the default\_text-based fields that we added to \irdatasets, except for ChatNoir that is a full-rank software for the ClueWeb corpora and uses different fields (title, body, etc.). The lexical approaches in PyTerrier and the dense approaches in BEIR can be configured as full-rank software (i.e., a first component building an index and a second component retrieving from the index) or re-rank software---but are just counted as one approach in Table~\ref{table-overview-retrieval-software}. All duoT5 and PyGaggle approaches only work as re-rankers. For ColBERT, we only use the re-rank variant, as ColBERT indices become very large.

In TIREx, all of these variants are available. To increase result comparability, however, our analysis fixes the first stage rankers to ChatNoir for the ClueWeb corpora and PyTerrier~BM25 on all other corpora. Their respective results are then handed to the total of 50~available  re-ranking approaches mentioned above. Altogether, 50~approaches are executed on all 32~tasks listed in Table~\ref{table-overview-tasks}. We executed the lexical approaches using 1~CPU and 10\,GB RAM, while all other approaches had additional access to a GeForce~GTX~1080 GPU with 8\,GB~RAM. Some models fail on this GPU as 8\,GB of~RAM do not suffice: ColBERT and two SBERT models failed on a few tasks, while the 3~billion parameter monoT5~/~duoT5 failed on all tasks. To handle these cases, we added two runners with access to an A100~GPU with 40\,GB~RAM to TIRA, which was sufficient. TIRA manages metadata about the resources used to produce a run, making hardware difference between evaluations transparent.

\begin{table}
\centering
\small
\renewcommand{\arraystretch}{1}%
\setlength{\tabcolsep}{1pt}%
\caption{Overview of the retrieval frameworks and the 50~retrieval approaches imported into TIREx.}
\label{table-overview-retrieval-software}
\vspace*{-2ex}
\begin{tabular}{@{}lll@{\kern-1em}cc@{}}
\toprule
\bfseries Framework             & \bfseries Type   & \bfseries Description & \multicolumn{2}{@{}c@{}}{\bfseries Approaches}     \\
\cmidrule{4-5}
                                &                  &                       &  Full-rank   &               Re-rank               \\
\midrule
BEIR~\cite{thakur:2021}         & Bi-encoder       & Dense retrieval       &      17      &                 17                  \\
ChatNoir~\cite{bevendorff:2018} & BM25F            & Elasticsearch cluster & \phantom{1}1 &            \phantom{1}0             \\
ColBERT@PT~\cite{khattab:2020}  & Late interaction & PyTerrier plugin      & \phantom{1}0 &            \phantom{1}1             \\
DuoT5@PT~\cite{pradeep:2021}    & Cross-encoder    & Pairwise transformer  & \phantom{1}0 &            \phantom{1}3             \\
PyGaggle~\cite{lin:2021}        & Cross-encoder    & Pointwise transformer & \phantom{1}0 &            \phantom{1}8             \\
PyTerrier~\cite{macdonald:2021} & Lexical          & Traditional baselines &      20      &                 20                  \\
\midrule
Pyserini$^{*}$~\cite{lin:2021}  & Lexical          & Traditional baselines & \phantom{1}4 &            \phantom{1}4             \\
\bottomrule
\multicolumn{5}{@{}l@{}}{\footnotesize ${}^*$Our import of Pyserini was not ready during the experiments but is now available.}
\end{tabular}
\end{table}

\begin{table*}
\centering
\small
\renewcommand{\arraystretch}{1}%
\setlength{\tabcolsep}{3.3pt}%
\caption{Effectiveness scores (nDCG@10) on 14~corpora (31~tasks; ClueWeb22B excluded as no judgments yet) for selected approaches and the best, median, and worst of each group (scores macro-averaged for corpora with multiple associated tasks).}
\label{table-retrieval-effectiveness}
\vspace*{-2ex}
\begin{tabular}{@{}lccccccccccccccccc@{}}
\toprule
\bfseries Corpus & \bfseries ChatNoir & \multicolumn{4}{@{}c@{}}{\bfseries Lexical} & \bfseries Late Int. & \multicolumn{4}{@{}c@{}}{\bfseries Bi-Encoder} & \multicolumn{3}{@{}c@{}}{\bfseries duoT5} & \multicolumn{4}{@{}c@{}}{\bfseries PyGaggle} \\
\cmidrule(l@{\tabcolsep}r@{\tabcolsep}){2-2}
\cmidrule(l@{\tabcolsep}r@{\tabcolsep}){3-6}
\cmidrule(l@{\tabcolsep}r@{\tabcolsep}){7-7}
\cmidrule(l@{\tabcolsep}r@{\tabcolsep}){8-11}
\cmidrule(l@{\tabcolsep}r@{\tabcolsep}){12-14}
\cmidrule(l@{\tabcolsep}){15-18}
                 &                    & BM25 &     Best      & Median &    Worst    &       ColBERT       & TAS-B &     Best      & Median &     Worst     & Base & Large &             3b             & MonoT5 &     Best      & Median &   Worst    \\
\midrule
Antique          &        ---         & 0.51 &     0.53      &  0.51  &    0.36     &        0.47         & 0.40  &     0.49      &  0.44  &     0.30      & 0.54 & 0.46  &            0.52            &  0.51  & \textbf{0.54} &  0.51  &    0.45    \\
Args.me          &        ---         & 0.43 & \textbf{0.57} &  0.43  &    0.14     &        0.26         & 0.17  &     0.33      &  0.24  &     0.13      & 0.33 & 0.29  &            0.29            &  0.30  &     0.39      &  0.34  &    0.27    \\
CORD-19          &        ---         & 0.28 &     0.64      &  0.55  &    0.21     &        0.58         & 0.50  & \textbf{0.70} &  0.60  &     0.50      & 0.66 & 0.61  &            0.66            &  0.69  &     0.69      &  0.63  &    0.55    \\
ClueWeb09        &        0.16        & 0.18 & \textbf{0.24} &  0.18  &    0.12     &        0.17         & 0.16  &     0.20      &  0.17  &     0.13      & 0.15 & 0.15  &            0.18            &  0.17  &     0.19      &  0.17  &    0.12    \\
ClueWeb12        &   \textbf{0.36}    & 0.24 &     0.27      &  0.25  &    0.14     &        0.23         & 0.25  &     0.28      &  0.26  &     0.23      & 0.33 & 0.30  &            0.35            &  0.26  &     0.28      &  0.26  &    0.23    \\
Cranfield        &        ---         & 0.01 &     0.01      &  0.01  &    0.01     &        0.01         & 0.01  &     0.01      &  0.01  &     0.00      & 0.01 & 0.01  &            0.01            &  0.01  &     0.01      &  0.01  &    0.01    \\
Disks4+5         &        ---         & 0.44 &     0.46      &  0.44  &    0.37     &        0.46         & 0.39  &     0.49      &  0.43  &     0.37      & 0.45 & 0.38  &            0.44            &  0.53  & \textbf{0.57} &  0.53  &    0.43    \\
GOV              &        ---         & 0.22 &     0.24      &  0.22  &    0.15     &        0.23         & 0.22  &     0.27      &  0.24  &     0.21      & 0.19 & 0.15  &            0.22            &  0.26  & \textbf{0.29} &  0.26  &    0.22    \\
GOV2             &        ---         & 0.47 &     0.49      &  0.44  &    0.25     &        0.45         & 0.34  &     0.46      &  0.42  &     0.34      & 0.47 & 0.43  &            0.48            &  0.48  & \textbf{0.51} &  0.48  &    0.41    \\
MS MARCO         &        ---         & 0.49 &     0.50      &  0.48  &    0.37     &        0.69         & 0.64  &     0.71      &  0.66  &     0.64      & 0.64 & 0.57  &            0.63            &  0.71  & \textbf{0.74} &  0.71  &    0.63    \\
MEDLINE          &        ---         & 0.34 & \textbf{0.42} &  0.27  &    0.18     &        0.25         & 0.14  &     0.26      &  0.21  &     0.14      & 0.34 & 0.32  &            0.36            &  0.25  &     0.35      &  0.27  &    0.24    \\
NFCorpus         &        ---         & 0.27 &     0.28      &  0.27  &    0.26     &        0.27         & 0.25  &     0.29      &  0.26  &     0.24      & 0.28 & 0.24  &            0.29            &  0.30  & \textbf{0.31} &  0.30  &    0.28    \\
Vaswani          &        ---         & 0.45 &     0.46      &  0.45  &    0.30     &        0.43         & 0.34  &     0.44      &  0.38  &     0.22      & 0.41 & 0.34  &            0.46            &  0.31  & \textbf{0.48} &  0.41  &    0.08    \\
WaPo             &        ---         & 0.38 &     0.39      &  0.37  &    0.24     &        0.43         & 0.34  &     0.43      &  0.37  &     0.33      & 0.40 & 0.28  &            0.40            &  0.45  & \textbf{0.49} &  0.45  &    0.40    \\
\midrule
Avg.             &        ---         & 0.34 &     0.39      &  0.35  &    0.22     &        0.35         & 0.30  &     0.38      &  0.33  &     0.27      & 0.37 & 0.32  &            0.38            &  0.37  & \textbf{0.42} &  0.38  &    0.31    \\
\bottomrule
\end{tabular}
\end{table*}

Table~\ref{table-retrieval-effectiveness} shows the aggregated evaluation results on 31~tasks (leaving out the ClueWeb22 as there are no judgments yet). We report the effectiveness as nDCG@10 (macro-averaged in case a corpus is associated with multiple tasks) for BM25, ColBERT, TAS-B, all three duoT5 variants, and monoT5 (in its default configuration with its default model) and the best, median, and worst approaches from the groups of 20~lexical, 17~bi-encoder, and 8~PyGaggle approaches. All deep learning models were trained on MS~MARCO and thus substantially improve upon the lexical models on MS~MARCO. However, on other corpora the deep learning models work in a zero-shot manner so that sometimes a lexical approach achieves the highest effectiveness (Args.me, ClueWeb09, and MEDLINE). Our results further show that BM25 is not always the best lexical ranker (e.g., on Args.me: 0.43 vs.\ 0.57). The effectiveness gap between the best and the worst model of a group can be substantial on some corpora (e.g., lexical models on Args.me: 0.14 vs.\ 0.57), while being negligible on others (e.g., lexical models on NFCorpus). The leaderboards of TIREx as aggregated in Table~\ref{table-retrieval-effectiveness} allow to easily select competitive baselines for very different tasks---often much easier than before.

\subsection{Case Study: Reproducibility Analysis}

As an example of a post-hoc analysis enabled by TIREx, we use \reproeval to analyze to which degree system preferences from the TREC Deep Learning~2019 task can be reproduced on other tasks. For each preference between approaches on TREC Deep Learning~2019 (e.g., monoT5 with an nDCG@10 of~0.71 compared to BM25's~0.48 induces a clear system preference), we set the approach with the lower effectiveness on TREC Deep Learning~2019 as the ``baseline'' in \reproeval and the other approach as the ``advanced system''. We study the reproducibility of the preferences on two dimensions~\cite{breuer:2020}:
\Ni
the effect ratio of the reproduction, and
\Nii
the delta relative improvement of the reproduction.
The effect ratio measures to which degree the advanced system is still better than the baseline on the different task (1~indicates a perfect reproducibility, values between~0 and~1 indicate reproducibility with diminished improvements on the different task, and 0~indicates failed reproducibility), while the delta relative improvement measures the relative effectiveness difference of the advanced system to the baseline (0~indicates perfect reproducibility, values between~-1 and~0 indicate an increased relative improvement of the advanced system, values between~0 and~1 indicate a smaller relative improvement, and 1~indicates failed reproducibility).

\begin{table}
\centering
\small
\renewcommand{\arraystretch}{1}%
\setlength{\tabcolsep}{3.4pt}%
\caption{Reproducibility of TREC DL~2019 system preferences on other tasks. Success rate in percent (effect ratio > 0; tasks ordered by success rate) and the 25\%, 50\%, and 75\% quantiles for the effect ratio and delta relative improvement.}
\label{table-reproducibility}
\vspace*{-2ex}
\begin{tabular}{@{}l@{}cccccccc@{}}
\toprule
\bfseries Task      & \bfseries Rank & \bfseries Succ. & \multicolumn{3}{@{}c@{}}{\bfseries Effect Ratio } & \multicolumn{3}{@{}c@{}}{\bfseries Delta Rel. Impr.} \\
\cmidrule(l@{\tabcolsep}r@{\tabcolsep}){4-6}
\cmidrule(l@{\tabcolsep}){7-9}
                         &                &                 &      25\%       &      50\%       &     75\%      &      25\%       & 50\% &            75\%             \\
\midrule
TREC~DL~2020             &  \phantom{0}1  &      88.1       & \phantom{-}0.68 & \phantom{-}0.90 &     1.11      &      -0.03      & 0.02 &            0.08             \\
Touch{\'e}~2020 (Task 2) &  \phantom{0}2  &      77.1       & \phantom{-}0.12 & \phantom{-}0.38 &     0.73      &      -0.09      & 0.04 &            0.17             \\
Web track~2004           &  \phantom{0}3  &      75.5       & \phantom{-}0.01 & \phantom{-}0.29 &     0.89      &      -0.07      & 0.10 &            0.31             \\
TREC-7                   &  \phantom{0}4  &      73.9       &      -0.03      & \phantom{-}0.31 &     1.11      &      -0.02      & 0.12 &            0.34             \\
Core 2018                &  \phantom{0}5  &      70.2       &      -0.05      & \phantom{-}0.24 &     0.90      &      -0.03      & 0.13 &            0.35             \\
NFCorpus                 &       10       &      66.4       &      -0.06      & \phantom{-}0.06 &     0.32      & \phantom{-}0.02 & 0.23 &            0.42             \\
Web track~2003           &       15       &      57.8       &      -0.14      & \phantom{-}0.04 &     0.23      &      -0.08      & 0.15 &            0.36             \\
Web track~2009           &       20       &      44.1       &      -0.40      &      -0.04      &     0.26      & \phantom{-}0.00 & 0.30 &            0.52             \\
Web track~2010           &       25       &      36.3       &      -0.49      &      -0.14      &     0.18      & \phantom{-}0.03 & 0.32 &            0.59             \\
Web track~2013           &       30       &      31.0       &      -0.43      &      -0.21      &     0.13      & \phantom{-}0.06 & 0.30 &            0.63             \\
\bottomrule
\end{tabular}
\end{table}

Table~\ref{table-reproducibility} shows the results of the preference reproducibility analysis. We report the ratio of system preferences with a successful reproduction (i.e., effect ratio > 0) and the 25\%, 50\%, and 75\%~quantiles for the effect ratio and the relative delta improvement. We order the tasks by the percentage of successfully reproduced preferences and show the top-5 tasks and every fifth lower ranked task. Not that surprising, the reproducibility on the very similar TREC Deep Learning~2020 is very good (88.1\%) but declines fast for other tasks (e.g., only 57.8\% for the Web track~2003 on rank~15). Analyzing the quantiles yields similar observations (e.g., 50\% of the system preferences have an almost perfect effect ratio of~0.90 or higher for TREC Deep Learning~2020, while the Web track~2003 on rank~15 has a median effect ratio of~0.04).

\section{Discussion}

\paragraph{Potential Impact of TIREx}
We believe that TIREx can have a substantial conceptual impact as we see no alternative to blinded retrieval evaluations in the future (given the practice of training~LLMs on basically all available ground truth for IR and NLP~tasks~\cite{chung:2022}). Additionally, the platform eases the organization of reproducible IR~experiments with software submissions. Shared task organizers can simply provide the well-documented open-source baselines from TIRA as starting points for the participants and can also use the baselines to ensure some more diverse judgment pools, especially for tasks that attract few participants. For shared tasks that run multiple years on different data, the organizers can automatically re-run all approaches submitted to previous editions to track progress. TIREx combines leaderboards with immutable software, promoting provenance of results, and enabling researchers and reviewers to identify and locally reproduce good baselines.

The submission platform TIRA proved robust after its complete redevelopment~\cite{froebe:2023b}: two NLP tasks used TIRA at SemEval~2023~\cite{froebe:2023d,kiesel:2023a} for which 71~of the 171~registered teams created 647~runs with software submissions. Our initial retrieval experiments with TIREx produced another 1,600~runs on standard corpora in less than a week, showing the platform to be robust and to have the potential for scaling up. When adopted by shared tasks and in individual IR~experiments, TIREx can become a (federated) hub for IR~resources and serve as a reference for reviewers. If a sufficient number of retrieval approaches, corpora, and supplementary data (e.g., manual query reformulations) are available through TIREx, integrating new resources gives direct access to an entire ecosystem, furthering the nascent standardization of IR~experiments.

\paragraph{Future Extensions of TIREx}
Interesting directions for future development besides including further IR~frameworks and libraries are integrations of TIREx with the IR~Anthology~\cite{potthast:2021n} and with Diff\-IR~\cite{jose:2021}. An integration with the IR~Anthology would enable links between entries in the TIREx leaderboards and the corresponding publications in the IR~Anthology to provide more detailed information on an approach but also to ``extend'' a publication by adding results on different corpora than originally used and putting an approach in a broader context with other approaches run on the same data. An integration with DiffIR would enable the rendering of runs as search engine result pages to easily contrast the quantitative evaluations already possible via the integrated \irmeasures with more qualitative evaluations of ranking differences or even (basic) user studies.

\section{Conclusion}

With TIREx---The IR Experiment Platform---we aim to substantially ease conducting (blinded) IR~experiments and organizing ``always-on'' reproducible shared tasks on the basis of software submissions. TIREx~integrates \irdatasets, \irmeasures, and PyTerrier with~TIRA. Retrieval workflows can be executed on-demand via cloud-native orchestration, reducing the effort for reproducing IR~experiments since software submitted to TIREx can be re-executed in post-hoc experiments. The platform has no lock-in effect, as archived experiments are fully self-contained, work stand-alone, and are easily exported. By keeping test data private, TIREx promotes further standardization and provenance of IR~experiments following the example of, e.g., medicine, where blinded experiments are the norm. TIREx is open to the IR~community and ready to include more corpora, shared tasks, and retrieval approaches.

\begin{acks}
This work has been partially supported by the OpenWebSearch.eu project (funded by the~EU; GA~101070014).
\end{acks}

\balance
\bibliographystyle{ACM-Reference-Format}
%\bibliography{sigir23-ir-experiment-platform-lit}

\end{document}